\documentclass[%
 reprint,
 amsmath,amssymb,
aps,
pra
]{revtex4-2}

\usepackage{comment}
\usepackage{graphicx}
\usepackage{dcolumn}
\usepackage[normalem]{ulem}
\usepackage{xcolor}
\newcommand{\dif}{\text{d}}
\newcommand{\tr}{\text{Tr}}

\newcommand{\norm}[1]{\left\lVert#1\right\rVert}
\usepackage{bm}
\begin{document}

\preprint{APS/123-QED}

\title{The concept of minimal dissipation and the identification of work in autonomous systems: A view from classical statistical physics}

\author{Anja Seegebrecht}
\email{anja.seegebrecht@physik.uni-freiburg.de}
\author{Tanja Schilling}%
\email{tanja.schilling@physik.uni-freiburg.de}

 \affiliation{Institute of Physics, University of Freiburg, Hermann-Herder-Stra\ss e 3, D-79104 Freiburg, Germany.}

\date{\today}

\begin{abstract}
Recently, the concept of minimal dissipation has been brought forward as a means to define work performed on open quantum systems [Phys.~Rev.~A {\bf 105}, 052216 (2022)]. We discuss this concept from the point of view of projection operator formalisms in classical statistical physics. We analyse an autonomous composite system which consists of a system and an environment in the most general sense (i.e.~we neither impose conditions on the coupling between system and environment nor on the properties of the environment). One condition any useful definition of work needs to fulfil is that it reproduces the thermodynamic notion of work in the limit of weak coupling to an environment that has infinite heat capacity. We propose a projection operator route to a definition of work that reaches this limit and we discuss its relation to minimal dissipation.
\end{abstract}

\maketitle

\section{\label{sec:intro}Introduction}

Work is a central concept both in thermodynamics and in mechanics. While the definition of the mechanical work done on an isolated, classical many-body system is straightforward, the notion of thermodynamic work is more involved. In thermodynamics, the definition of work requires a distinction between a system and its environment. Further, the properties of the environment need to be specified as well as conditions on the strength of the coupling to the system \cite{Balian_1991, Baumer_2018, Funo_2018, Brenig_2013}. If the coupling is strong, the system and the environment are correlated, or the environment contains only few degrees of freedom, it is unclear how to define work and whether it is a meaningful concept at all. \footnote{This statement holds for classical as well as quantum mechanical systems.}

One requirement a definition of work needs to fulfil is that in the limit of a quasi-static process performed on a system coupled to a heat bath it reproduces the definition given in equilibrium thermodynamics. (In this article we use the term {\it heat bath} for an environment with infinite heat capacity, which is weakly coupled to the system of interest, while we use the term {\it environment} for any type of system coupled in any way to the system of interest.) I.e.~if work is done on a system by externally changing some of its parameters (such as the strength of a magnetic field or the volume available to the system) and if these parameters are varied infinitely slowly and the system is in contact with a heat bath, the total work is the difference between the final and the initial equilibrium free energy of the system: $W = \Delta F= F_f-F_i$. \cite{Landau_1981, Brenig_2013}
In such a quasi-static transformation the system remains in equilibrium with the bath and the total entropy does not change \cite{Balian_1991}.
If the parameters are changed faster, the process is in general irreversible and $W$ will, on average, exceed the free energy difference $\langle W \rangle \geq \Delta F$. \cite{Landau_1981}
Or conversely, if we wish to extract work from the system, it will be less than the available free energy. 

A variety of definitions of the work done on open quantum systems have been published and discussed \cite{Pusz_1978, Alicki_1979, Kosloff_2013, Dann_2023, Strasberg_2017, Baumer_2018, Funo_2018, Rivas_2020, Seegebrecht_2024}, but so far no consensus has been reached. In this article we will analyse the concept of minimal dissipation, which has recently been brought forward as a means to obtain a definition of work \cite{Colla_2021}. In particular, we will discuss the relation between the free energy and the effective Hamiltonian obtained by the condition of minimal dissipation.

We will use the framework of autonomous systems. In this framework, instead of imposing an external driving force on the Hamiltonian or coupling the system to a specific type of bath, one considers a composite, isolated supersystem made of the system $S$ and its environment $E$. This composite is governed by a Hamiltonian that is independent of time. In such a setting, the environment $E$ acts as a source of work for the system $S$, i.e.~the time-dependent external driving, which is required for work to be performed on $S$, is produced by the dynamics of $E$. This is a standard approach used for example in ref.~\cite{Silva_2021, Weimer_2008, Barra_2015, Valente_2018, Rodrigues_2019, Malavazi_2022, Colla_2021, Alipour_2022, Ahmadi_2023, Alipour_2016}.

We obtain the dynamics of $S$ by tracing out the degrees of freedom of $E$. 
The resulting equation of motion (EoM) for the density operator of $S$ can in general be written in the form 
\begin{equation}
\dot{\rho}_S(t)=-i[H^{\rm eff}_S(t),\rho_S(t)] + \mathcal{D}_t[\rho_S(t)] \; ,
\end{equation}
i.e.~in terms of a commutator that contains an effective, time-dependent Hamiltonian and a rest. In the case of a classical system, the density operator is replaced by the phase space probability density and the commutator by a Poisson bracket, but the structure of the equation remains the same.
In the literature the first term is often called the {\it conservative} part and the rest the {\it dissipative} part \cite{Agarwal_1973, Colla_2021, Izvekov_2021, hijon2010} (we will see later that these names can be misleading, as the interpretation in terms of dissipation is not always given). Work is then defined via the equation for the work flux related to the effective Hamiltonian \begin{equation}
    \dot{W}_S(t) = \text{Tr} \Big(\dot{H}_S^{\rm eff} \, \rho_S\Big) \; .
\end{equation} 

To systema\-tically integrate out degrees of freedom of an autonomous system has the advantage over other approaches, that the setting is as general as possible, i.e.~a priori neither assumptions on the coupling between the system and the environment nor on the properties of the environment are required. Several authors have brought forward definitions of work based on the effective Hamiltonian in such a setting \cite{Weimer_2008, Alipour_2016, Alipour_2022, Ahmadi_2023, Colla_2021, Picatoste_2024}. However, as the splitting between the conservative part and the dissipative part is not unique, different authors have suggested different identifications of internal energy, work and heat \cite{Seegebrecht_2024}. Here we will analyse which type of splitting produces an effective Hamiltonian that equals the Hamiltonian of mean force in the limit of a quasi-static process.

We will begin with an introduction to our notation that highlights the similarity between the classical and the quantum mechanical description of the dynamics. We proceed with the discussion of projection operator techniques applied to classical supersystems. If one uses the projection operator introduced by Zwanzig \cite{zwanzig2001} to integrate out the degrees of freedom of the environment then, in the classical case, one obtains an EoM for the system which contains the Hamiltonian of mean force in the conservative part, i.e.~in a Poisson bracket \cite{Izvekov_2021}. We briefly recall Izvekov's derivation of this equation. Then we show that the EoM for the corresponding relevant density has a similar structure. We argue that this EoM is equal to the one obtained via the principle of minimal dissipation, however this requires a different choice of inner product than used in the original work by Colla and Breuer \cite{Colla_2021}.
Finally, we show that this line of reasoning cannot be extended to open quantum systems in a general manner but that it does hold for systems and environments with equilibrium states that factorize.

\section{Classical-quantum analogy}

When studying the classical case, we work with an ensemble of systems, each of which has a state $\Gamma$ from a phase space $\Phi$ and a Hamiltonian $H$. These systems are distributed according to a phase space probability density $\rho(\Gamma, t)$. When we discuss quantum mechanical systems, we also use the symbol $\rho$, but then we mean the density matrix (also called the statistical operator) which acts in a Hilbert space $\mathcal{H}$.
The density is positive and normalized, i.e.~
\begin{eqnarray}
    \rho &\geq& 0 \quad \Leftrightarrow \begin{cases} \rho(\Gamma) \geq 0, \ \forall \Gamma \in \Phi \\ \langle \psi \vert \rho \vert \psi \rangle \geq 0 , \ \forall \vert \psi \rangle \in \mathcal{H}   \end{cases} \\
    \tr(\rho) &=& 1 \quad \Leftrightarrow \begin{cases} \int \dif \Gamma \rho(\Gamma) =1 \\ \sum_i \langle \varphi_i \vert \rho \vert \varphi_i \rangle = 1    \end{cases}
\end{eqnarray}
where the first line of each equation refers to the classical case and the second line to the quantum mechanical case.
To highlight the structural similarities we use the symbol $\tr( \,\cdot{}\,)$ in both cases and understand it either as the integration over all phase space points or as the summation over matrix elements with a complete orthonormal basis $\{\varphi_i\}$ of the Hilbert space $\mathcal{H}$.

The evolution of the microscopic state is determined by the Liouville equation
\begin{equation}
    \dot{\rho}(t) = -i \mathcal{L}\rho (t) \label{eq:Liouville_equation}
\end{equation}
where the Liouvillian either acts on a phase space function $X=X(\Gamma)$ as the Poisson bracket or on a Hilbert space operator $X$ as the commutator:
\begin{equation}
    i\mathcal{L} X = \begin{cases} -\{H,X\} = \mathbb{J}\nabla_\Gamma H \cdot \nabla_\Gamma X   \\ \frac{i}{\hbar} [H,X] = \frac{i}{\hbar} (HX-XH)   \end{cases}
\end{equation}
 Here we introduced the symplectic matrix  $\mathbb{J}=\begin{pmatrix} 0 & \mathbb{I} \\ -\mathbb{I} & 0 \end{pmatrix}$, where $\mathbb{I}$ is the identity matrix of the system. From now on we will use $\hbar =1$. 
 We are interested in autonomous systems which are governed by time-independent Hamiltonians and thus will restrict the discussion to time-independent Liouvillians here.

The Liouville-equation, eq.~\eqref{eq:Liouville_equation}, has the formal solution
\begin{equation}
    \rho (t) = e^{-i\mathcal{L}t} \rho(0).
\end{equation}

An observable $B$ is represented by a phase space function or self-adjoint operator, respectively. We consider observables that are not explicitly time-dependent. Expectation values at time $t$ are given by 
\begin{equation}
    \langle B \rangle_t = \tr(B \rho(t)) = \tr(B e^{-i\mathcal{L}t} \rho(0)) .
\end{equation}

This expression is formulated in the Schrödinger picture, i.e., the time-dependence is carried by the density. Equivalently the Heisenberg picture can be used, where the time-dependence is carried by the observables and the average is taken with respect to the initial density $\langle B\rangle_t= \tr(\rho(0) B_H(t))$. Since these expectation values are the same, the Liouville equation for the observable $B_H(t)$ is
\begin{eqnarray}
    \frac{\dif}{\dif t}B_H(t) = i\mathcal{L}_H B_H(t) = e^{i\mathcal{L}t}i\mathcal{L}B \label{eq:HeisenbergEoM}\\
    \text{ with } i\mathcal{L}_H X= \begin{cases} -\{H_H,X\} \\ i[H_H,X]  \end{cases}. \label{eq:obs_Liouville}
\end{eqnarray}

\subsection{Hamiltonian of mean force}
\label{sec:HMF}
We are interested in the description of a subsystem of a composite system, i.e., only the degrees of freedom of the subsystem are of interest and the rest is treated as the environment.
The Hamiltonian of mean force is a useful analytical tool to treat the average effect of environmental degrees of freedom on variables of interest \cite{MacKerell_2001}. We briefly recall its definition:
The Hamiltonian of a bipartite system has the general form $H=H_S+H_E+H_{SE}$, where $H_S$ acts only on the subsystem $S$ and $H_E$ acts only on the environment $E$. The third term $H_{SE}$ represents the interaction operator.
A thermal equilibrium state at some inverse temperature $\beta$ is given by $\varrho_\beta = e^{-\beta H}/Z$ with the partition function $Z=\tr(e^{-\beta H})$. (Note that we do not imply here that the composite system was coupled to some additional external heat bath, we simply consider an isolated supersystem prepared in a state of the canonical form.) 
Then the reduced state of the system $S$ is found by tracing over the environmental degrees of freedom
\begin{eqnarray}
    \varrho_{S,\beta} := \tr_E(\varrho_\beta) = e^{-\beta H^\ast_S}/Z^\ast \; .\label{eq:reduced_equilibrium}
\end{eqnarray}
with the Hamiltonian of mean force
\begin{equation} 
    H^\ast_S := -k_B T \ln \tr_E(e^{-\beta H})/\tr_E(e^{-\beta H_E}) \label{eq:H_mean_force} \; .
\end{equation}

In general, $H^\ast_S$ differs from the bare system Hamiltonian $H_S$ \cite{Trushechkin_2022}. Eq.~\eqref{eq:reduced_equilibrium} only determines $H_S^\ast$ up to an additive constant. Any addition would also change the partition function and leave averages unaffected. 
The expression~\eqref{eq:H_mean_force} is based on the common choice $Z^\ast_S = Z/Z_E$ \cite{Trushechkin_2022, Roux_2001}. 

The term {\it Hamiltonian of mean force} is used predominantly in the context of quantum mechanics. In physical chemistry, the same quantity is called {\it effective free energy} or {\it restricted free energy} \cite{schilling2022, Izvekov_2021}. In the context of molecular modelling of biomolecules and polymers a related quantity is often used, the {\it potential of mean force}. The potential of mean force is obtained by tracing not only over the environment $E$ in eq.~\eqref{eq:H_mean_force} but also over the momenta of the particles in $S$, i.e.~over the contribution of the kinetic energy of the system.

The relation between the Hamiltonian of mean force and the thermodynamic work is made as follows:
The free energy of $S$ is obtained by taking the logarithm of the partition function associated with the Hamiltonian of mean force
\begin{equation}
    F_S = -k_BT \ln Z_S^\ast = -k_BT \ln \tr_S (e^{-\beta H_S^\ast}) \; , \label{eq:free_energy}
\end{equation}
i.e.~the Hamiltonian of mean force is directly related to the free energy. If the energy contained in the coupling term $H_{SE}$ is small compared to $H_S$ and if the environment is sufficiently large to stay in a canonical state, then the free energy difference equals the work performed in a quasi-static process.

\section{Projection operator technique} 

Projection operator techniques can be applied to derive EoM for macroscopic variables or for the degrees of freedom of the subsystem of interest by decomposing the dynamics. The first step is the definition of a projector that maps the space of relevant variables on itself.  
Such an operator can be defined by the choice of an appropriate scalar product or a relevant density \cite{Grabert_1977_equivalence}. 

We first treat the classical case and follow ref.~\cite{Izvekov_2021}. Let $A=\{A_k \}_k$ be a set of relevant independent observables. (In the context of projection operator formalisms the term {\it relevant} is commonly used for degrees of freedom or observables that are not integrated out. \cite{zwanzig2001,Grabert_1982} This does not imply that the formalism only works if the other degrees of freedom are less relevant. The derivations hold in any case, the naming convention is just somewhat misleading.) We define $O_A \subset O$ as the space of observables which are fully determined through $A$ (in the classical system $B \in O_A$ implies $B(\Gamma)=B(A(\Gamma))$ only depends on $\Gamma$ through $A$). With the help of a projection operator we can decompose each $B \in O$ into a component in $O_A$ and a component in the orthogonal space. 

Now we define a projection operator acting on the space of all observables with image in $O_A$. This projection operator may be represented by
\begin{equation}
    \mathcal{P}B = \sum_{k,l} (B,\phi_l)(\phi_l,\phi_k)^{-1} \phi_k \label{eq:general_projection}
\end{equation}
where the set $\{\phi_k\}_k $ forms a possibly incomplete basis of $O_A$. For convenience, usually $(\phi_k,\phi_l)=\delta_{kl}$ is chosen. This simplifies eq.~\eqref{eq:general_projection} to $\sum_k (B,\phi_k) \phi_k$. Eq.~\eqref{eq:general_projection} clearly defines an idempotent map ($\mathcal{P}\mathcal{P}=\mathcal{P}$). It projects out the $\phi_k$ and is linear. 

Complementary to the projector we can define the map $\mathcal{Q}=(\mathbb{I}-\mathcal{P})$ and decompose the dynamics into a relevant part $\mathcal{L}_AB \in O_A$ and a contribution in the space orthogonal to $O_A$. 
(In general, we could deal with time-dependent projectors by means of a time-dependent basis.)

With such a projector we can split the Liouville equation for any observable into a part determined by the relevant observables and a part that stems from the degrees of freedom that have been integrated out, i.e.~from the space orthogonal to the relevant observables. We will later define a projection operator that will allow us to identify the former with the conservative part and the latter with the dissipative part of a master equation.  

To carry out the splitting, we use the identity \cite{Izvekov_2021, Grabert_1982} 
\begin{equation}\begin{split}
    e^{i\mathcal{L}t} &= e^{i\mathcal{L}t} \mathcal{P} + \int_0^t \dif s e^{i\mathcal{L}s} \mathcal{P}i\mathcal{L}\mathcal{Q} G(s,t)\\
    &+ \mathcal{Q}G(0,t) \label{eq:identity_heisenberg}
\end{split}
\end{equation}
with the propagator for the orthogonal dynamics
\begin{equation}
    G(s,t) = e^{i\mathcal{L}\mathcal{Q}(t-s)}.
\end{equation}
(See apx.~\ref{apx:decomposition} for details on the derivation.)

With eq.~\eqref{eq:identity_heisenberg} and the EoM for observables, eq.~\eqref{eq:HeisenbergEoM}, we obtain the time-convolution equation 
\begin{equation}\begin{split}
    \frac{\dif}{\dif t} B_H(t) &= e^{i\mathcal{L}t}\mathcal{P} i\mathcal{L} B + \int_0^t \dif s e^{i\mathcal{L}s} \mathcal{P}i\mathcal{L} \mathcal{Q} e^{i\mathcal{L}\mathcal{Q}(t-s)} i\mathcal{L}B\\ &+ \mathcal{Q} e^{i\mathcal{L}\mathcal{Q}t} i\mathcal{L}B \label{eq:obs_EoM}.
    \end{split}
\end{equation}

Alternatively, we can construct an EoM for the so-called {\it relevant density}, i.e.~the probability density associated with the relevant observables. Then we work with the adjoint projector defined through
\begin{equation}
    \tr(\mu \mathcal{P}X)=\tr(X\mathcal{P}^\dagger \mu) \label{eq:adjoint_projector}
\end{equation}
acting on a probability density $\mu$. Any scalar product can be related to the Hilbert-Schmidt product $(X,Y)_{HS} = \tr(X^\dagger Y)$ by defining a transformation $\Sigma$ such that $(X,Y) = \tr((\Sigma X)^\dagger Y)$ \cite{Grabert_1977_equivalence}.
Thus,
\begin{eqnarray}
    \tr(\mu \mathcal{P} X ) &=& \sum_k \tr(\mu (X,\phi_k) \phi_k) \\
    &=& \sum_k \tr(\mu \tr(X \Sigma \phi_k)\phi_k ) \\
    &=& \sum_k \tr(X \Sigma \phi_k \tr(\mu \phi_k)) \; .
\end{eqnarray}
Applied to the probability of the composite system $\rho$ the adjoint projector yields the relevant density $\sigma:=\mathcal{P}^\dagger \rho= \sum_k \tr(\rho \phi_k) \Sigma \phi_k$.

We then decompose the f-propagator (the propagator in the Schrödinger representation) 
\begin{eqnarray}
    e^{-i\mathcal{L}t} &=& \mathcal{P}^\dagger e^{-i\mathcal{L}t} + G^\dagger(t,0) \mathcal{Q}^\dagger  \\
    & & - \int_0^t \dif s  G^\dagger(t,s) Q^\dagger i \mathcal{L} \mathcal{P}^\dagger e^{-i\mathcal{L}s} \; ,
\end{eqnarray}
and thus obtain
\begin{eqnarray}
    \dot{\rho} (t) &=& -i\mathcal{L}e^{-i\mathcal{L}t}\rho(0) \\
    &=& - i\mathcal{L}\mathcal{P}^\dagger \rho(t) -\int_0^t \dif s \mathcal{L} e^{-i \mathcal{Q}^\dagger \mathcal{L}(t-s)} \mathcal{Q}^\dagger \mathcal{L} \mathcal{P}^\dagger \rho(s) \nonumber \\
    & &-i \mathcal{L} e^{-i\mathcal{Q}^\dagger \mathcal{L}t}\mathcal{Q}^\dagger \rho(0) \; .\label{eq:density_EoM}
\end{eqnarray}

 Now the task is to define a projection operator that will turn the first term in eq.~\eqref{eq:obs_EoM} or in eq.~\eqref{eq:density_EoM} into a Poisson bracket containing a Hamiltonian of mean force.

\section{Zwanzig projector in classical statistical mechanics} 
\label{sec:Zwanzig}

To define the Zwanzig projector, we first introduce functions $\psi_\alpha$ which fix the values of the relevant observables $A_k$ to the numerical values $\alpha_k$.
\begin{align}
    \psi_\alpha (\Gamma) := \delta (A(\Gamma)-\alpha) = \prod_k \delta(A_k(\Gamma)-\alpha_k)  \;. 
\end{align}
These form a set of functions with the continuous index $\alpha$ and with the useful property 
\begin{equation}
    \psi_\alpha \psi_{\alpha'} = \delta (\alpha-\alpha') \psi_\alpha\;. \label{eq:classical_projections_psi}
\end{equation}

The Zwanzig projector in the Heisenberg picture is given by 
\begin{eqnarray}
    \mathcal{P}B(\Gamma) = \frac{\tr(\varrho \psi_{A(\Gamma)} B)}{\tr(\varrho \psi_{A(\Gamma)})} = \int \dif \alpha \frac{\tr(\varrho \psi_\alpha B)}{\tr (\varrho \psi_\alpha)} \psi_\alpha(\Gamma) \label{eq:defZwanzig}
\end{eqnarray}
with some probability distribution $\varrho$ \cite{Zwanzig_1961, Grabert_1982}. (We will later set $\varrho$ equal to the canonical distribution in order to obtain a thermodynamic interpretation of certain terms in the EoM. For the moment, however, we work with the general case.)

The value $p(\alpha,t)=\tr(\rho(t)\psi_\alpha)$ defines the so-called "macroscopic probability density" of the observables $A$, i.e.~in an ensemble that is distributed according to $\varrho$, $p(\alpha,t) \dif \alpha$ is the probability to find the values of the observables $A$ in the volume element $\dif \alpha$ around the values $\alpha$.

The trace in the numerator of the second term of eq.~\eqref{eq:defZwanzig} integrates $B(\Gamma)$ over all microstates $\Gamma$ for which $A(\Gamma)=\alpha$, where the microstates are weighted according to the probability distribution $\varrho(\Gamma)$. Hence, the Zwanzig projector contains a conditional probability in the ensemble specified by $\varrho(\Gamma)$. It maps the observable $B$ to the best possible approximation of $B$ in terms of functions of $A$ \cite{Chorin_2000}. This is the property we need in order to define a Hamiltonian of mean force.

The connection of eq.~\eqref{eq:defZwanzig} to the projection operator, eq.~\eqref{eq:general_projection}, becomes clear if we identify the scalar product 
\begin{equation}
    (X,Y) = \tr(\varrho XY) \label{eq:classical_scalar_product}
\end{equation} and replace the sum over $k$ by the integral over the state space. The corresponding transformation is simply $\Sigma X = \varrho X$.

 The adjoint projector defined by eq.~\eqref{eq:adjoint_projector} is given by
 \begin{equation}
     \mathcal{P}^\dagger \mu(\Gamma) = \varrho(\Gamma)  \int \dif \alpha \frac{\tr(\psi_\alpha \mu) }{\tr(\psi_\alpha \varrho)} \psi_\alpha(\Gamma) \; .
 \end{equation}
 We clearly have the relation
 \begin{equation}
     \mathcal{P}^\dagger (\varrho X) = \varrho \mathcal{P}X \; .
 \end{equation}

While $\mathcal{P}$ projects out the $\psi_\alpha$, $\mathcal{P}^\dagger$ projects out the relevant density $\sigma$, which yields the same macroscopic probability density as the density of the composite supersystem $\rho$ 
\begin{equation}
    p(\alpha,t)=\tr(\psi_\alpha \sigma) \; .
\end{equation}

\subsection{Dirft Term, Conservative Force}
\label{sec:EoMDrift}
 If we apply the EoM for observables, eq.~\eqref{eq:obs_EoM}, to the relevant observables themselves, the first term can be written as
\begin{eqnarray}
    e^{i\mathcal{L}t} \mathcal{P}i\mathcal{L}A &=& -\int \dif \alpha \frac{\tr( \varrho \psi_\alpha \{H,A\})}{\tr (\varrho \psi_\alpha)} e^{i\mathcal{L}t} \psi_{\alpha }\; . \label{eq:EoMrelevantObsFirstTerm}
\end{eqnarray}

Depending on the context, in the literature on classical systems this term is sometimes called "drift" \cite{Grabert_1982, schilling2022} and sometimes "conservative force" \cite{rudzinski2019}. 
We now set the weight in the projector to the canonical equilibrium distribution $\varrho= \varrho_\beta (\Gamma)= e^{-\beta H(\Gamma)}/\tr(e^{-\beta H})$. (Note that this does not imply that the system were coupled to a heat bath. We simply choose a weight to define a specific projection operator, but we still consider the equations of motion for the general case.)
Then we can exploit the fact that for any observable, and particularly $A$:
\begin{eqnarray}
    i\mathcal{L}(\varrho_\beta A) &=& i\mathcal{L}(\varrho_\beta)A + \varrho_\beta i\mathcal{L}(A) \nonumber \\
    &=& - \beta \varrho_\beta \{H,H\} A - \varrho_\beta \{H,A\} \nonumber \\
    &=& \varrho_\beta i\mathcal{L}A. \label{eq:detailed_balance}
\end{eqnarray}

and 
\begin{equation}
    \varrho_\beta \{H,A\} = - k_BT \{\varrho_\beta, A\} \label{eq:Poisson_relation}
\end{equation}
If $A$ is a subset of a set of canonical variables (e.g., all positions and momenta of the particles in $S$) we can use $\{A,\, \cdot{ } \}_\Gamma = \{A, \, \cdot \, \}_A =- \mathbb{J}\nabla_A \, \cdot \,$.

The numerator on the right of eq.~\eqref{eq:EoMrelevantObsFirstTerm} becomes $-k_BT\tr(\psi_\alpha \{A,\rho_\beta\})$.  
The propagator only acts on the $\psi_\alpha(\Gamma)$ and yields $\exp(i\mathcal{L}t) f(A) = f(A_H(t))$. 
Thus,
\begin{widetext}

\begin{eqnarray}
    e^{i\mathcal{L}t} \mathcal{P}i\mathcal{L}A &=& -k_BT \int \dif \alpha \frac{\mathbb{J}\nabla_A \tr(\varrho_\beta \psi_\alpha)\vert_{A(\Gamma)=\alpha}}{\tr(\varrho_\beta \psi_\alpha)} \delta(A_H(t)-\alpha) \\
    &=& -k_BT \int \dif \alpha \mathbb{J}\nabla_A \ln\tr(\varrho_\beta \psi_\alpha)\vert_{A(\Gamma)=\alpha} \delta(A_H(t)-\alpha) \\
    &=& -k_B t \mathbb{J} \nabla_A \ln \int \dif \Gamma \varrho_\beta (\Gamma) \delta (A(\Gamma)  - A (\Gamma(t)) \\
    &=:& -k_BT \mathbb{J}\nabla_A \ln \tr(\varrho_\beta \psi_{A_H(t)}) \\
    &=:& \{A,H^\ast(A,t)\}.
\end{eqnarray}
\end{widetext}
In the last step, we have identified the Hamiltonian of mean force 
associated with the values of $A$ taken in the equilibrium ensemble $H^\ast(A,t)= - k_BT \ln \tr(\varrho_\beta \psi_{A_H(t)})$. As shown in sec.~\ref{sec:HMF}, the free energy is obtained by tracing over $H^\ast(A,t)$. 
So by choosing the appropriate projection operator it is indeed possible to construct an EoM, in which the effective Hamiltonian that appears in the Poisson bracket has a direct connection to the thermodynamic potential and can be used to determine work done by the mean forces. 
If the environment remains in a canonical state and evolves slowly, then both the temperature and the mean force applied to the subsystem would change quasistatically. 

If we set the observables to be positions and momenta of the system $A=\Gamma_S=(q^1,\dots, q^n, p^1, \dots p^n)$ the Poisson bracket $\{H,\Gamma_S\}_\Gamma = \{H, \Gamma_S\}_{\Gamma_S}$ since $\frac{\partial}{\partial q^m} \Gamma_S = 0 = \frac{\partial}{\partial p^m} \Gamma_S $ for $m>n$. 
Now, eq.~\eqref{eq:EoMrelevantObsFirstTerm} can be rewritten
\begin{widetext}
\begin{eqnarray}
    e^{i\mathcal{L}t}\mathcal{P}i\mathcal{L} \Gamma_S &=& k_B T\frac{\int \dif \Gamma  \delta (\Gamma_S-\Gamma_S(t))\{\varrho_\beta (\Gamma),\Gamma_S\}_\Gamma }{\int \dif \Gamma \varrho_\beta(\Gamma) \delta (\Gamma_S-\Gamma_S(t))} \\
    &=& -k_BT \mathbb{J} \frac{\nabla_{\Gamma_S} \int \dif \Gamma \delta (\Gamma_S- \Gamma_S(t)) \varrho_\beta(\Gamma)}{ \tr_E(\varrho_\beta)\vert_{\Gamma_S=\Gamma_S(t)}} \\
    &=&  -k_BT \mathbb{J} \nabla_{\Gamma_S} \ln \tr_E(\varrho_\beta)\vert_{\Gamma_S=\Gamma_S(t)} \\
    &=& \{\Gamma_S, H^\ast(\Gamma_S,t)\}_{\Gamma_S} = \{\Gamma_S, H^\ast(\Gamma_S,t) \}_\Gamma
\end{eqnarray}
\end{widetext}

In summary, so far we have integrated out the degrees of freedom of the environment without making any approximations, and we have obtained an equation of motion for the system, which contains a Poisson bracket with the Hamiltonian of mean force, i.e.~with the quantity that is directly related to the thermodynamic equilibrium work  
\cite{Izvekov_2021}. 

In the context of open quantum systems, one usually starts out from the EoM for the density matrix rather than the EoM for the observables.
Hence, in analogy to the derivation just presented, we now analyse the first term in eq.~\eqref{eq:density_EoM}. Since we are interested only in the evolution of the relevant observables which are described by $p(\alpha,t)$, we multiply by $\psi_\alpha$ and take the trace:
\begin{widetext}

\begin{eqnarray}
    -i\tr(\psi_\alpha \mathcal{L} \mathcal{P}^\dagger \rho(t))
    &=&   -i\tr \left(\psi_{\alpha} \varrho  \int \dif \alpha' \frac{\tr(\psi_{\alpha'} \rho(t))}{\tr(\psi_{\alpha'} \varrho)} \mathcal{L}\psi_{\alpha'}\right) \\
    &=& -i\int \dif \alpha' \frac{\tr(\psi_{\alpha'} \rho(t))}{\tr(\psi_{\alpha'} \varrho)}\tr\left(   \psi_{\alpha'}\mathcal{L}\psi_{\alpha} \varrho \right)
\end{eqnarray}   
\end{widetext}
In the last step we exploit that, $i\mathcal{L}=(i\mathcal{L})^\dagger$. 
If $\varrho=\varrho_\beta$ we can use eq.~\eqref{eq:detailed_balance} again. The Liouville operator is a first order differential operator in phase space and we can use the chain rule \cite{Zwanzig_1961}

\begin{eqnarray}
    -i\mathcal{L}\psi_\alpha 
    &=& - \nabla_A \delta (A-\alpha) \cdot i\mathcal{L}A =\nabla_\alpha \cdot \psi_\alpha i\mathcal{L}A.
\end{eqnarray}
With $\psi_\alpha \psi_{\alpha'} =  \delta (\alpha-\alpha')\psi_\alpha$ we get
\begin{widetext}
\begin{eqnarray}
\label{eq:densityDriftHamMeanForce}
    -i\tr(\psi_\alpha \mathcal{L} \mathcal{P}^\dagger \rho(t))
    &=& \nabla_\alpha \cdot \int \dif \alpha' \frac{\tr(\psi_{\alpha'} \rho(t))}{\tr(\psi_{\alpha'} \varrho_\beta)}\tr\left(   \psi_{\alpha} i\mathcal{L}A \varrho_\beta \right)\delta(\alpha-\alpha') \\
    &=& -\nabla_\alpha \cdot p(\alpha, t) \frac{\tr(\varrho_\beta \psi_\alpha \{H,A\})}{\tr(\varrho_\beta \psi_\alpha)} \\
    &=& - k_BT \nabla_\alpha\cdot p(\alpha,t) \frac{\mathbb{J} \nabla_\alpha \tr( \varrho_\beta \psi_\alpha)}{\tr( \varrho_\beta \psi_\alpha)} \\
    &=& \nabla_\alpha p(\alpha,t) \cdot \mathbb{J} \nabla_\alpha H^\ast(\alpha) \\
    &=& \{p(\alpha,t),H^\ast(\alpha)\}_\alpha 
\end{eqnarray}    
\end{widetext}
The fraction in the second line also appears in eq.~\eqref{eq:EoMrelevantObsFirstTerm}. It describes the average rate of change of the relevant observables $A$ in the conditional equilibrium ensemble \cite{zwanzig2001}.  

In both cases this fraction was turned into the Hamiltonian of mean force 
in the following steps. Accordingly, the drift term in the EoM for $p(\alpha,t)$ also features a Poisson bracket structure with $H^\ast(\alpha)$. (Note, it is crucial that $A$ represents a canonical set to obtain the symplectic structure.)

\section{Relation to Minimal Dissipation}
For the description of open quantum systems time-convolutionless master equations (TCL) of the form
\begin{equation}
\label{eq:TCL}
\dot{\rho}_S(t)=-i[H^{\rm eff}_S(t),\rho_S(t)] + \mathcal{D}_t[\rho_S(t)] \; ,
\end{equation}
 are often used, where $H^{\rm eff}_S$ is the effective Hamiltonian, $\mathcal{D}_t$ is a dissipator of the generalized Lindblad form and $\rho_S$ is the density matrix of the system.  The function $p(\alpha,t)$ defined in sec.~\ref{sec:Zwanzig} is the classical equivalent of $\rho_S$ if the observables $A$, which are set to the values $\alpha$ in $p(\alpha,t)$, are the canonical degrees of freedom of the system \cite{Shibata_1977}. 
In the form of eq.~\eqref{eq:TCL} the decomposition of the master equation in a conservative and a dissipative part is not unique. A distinct splitting can be achieved by generalizing the Hilbert-Schmidt scalar product for operators to superoperators and minimizing the dissipator with respect to the induced norm \cite{Hayden_2021}.  The resulting effective Hamiltonian is used to define work in ref.~\cite{Colla_2021, Picatoste_2024, Gatto_2024}.

In ref.~\cite{Colla_2025} it is pointed out that the effective Hamiltonian obtained in the minimal dissipation framework does not relax to the Hamiltonian of mean force in equilibrium.  We compare ref.~\cite{Colla_2021} with sec.~\ref{sec:EoMDrift} to identify the origin of this discrepancy. Our derivations differ in four points from the ones presented in ref.~\cite{Colla_2021}: a) eq.~\eqref{eq:density_EoM} is non-local in time, while eq.~\eqref{eq:TCL} is time-local, b) the definitions of the inner product differ, c) we did not explicitly impose the condition of minimal dissipation and d) in sec.~\ref{sec:EoMDrift} we focussed on the classical case, instead of quantum systems.

Interestingly, the non-locality in time is not the cause of the discrepancy. The operations that render eq.~\eqref{eq:TCL} time-convolutionless affect only the second term of eq.~\eqref{eq:density_EoM} and not the drift term. As shown by Los \cite{Los_2020}, one can remove the time-convolution from eq.~\eqref{eq:density_EoM} without affecting the drift. Hence the considerations discussed above also apply to eq.~\eqref{eq:TCL}.

The condition of minimal dissipation is not the source of the discrepancy, either. We did not impose the condition explicitly, however, our derivation fulfils it by construction. The splitting between the conservative term and the dissipative term in eq.~\eqref{eq:TCL} as well as in eq.~\eqref{eq:density_EoM} is determined by the projection operator. Depending on the choice of the functions $\{\phi_k\}$ in eq.~\eqref{eq:general_projection} contributions to the dynamics get shuffled from one part of eq.~\eqref{eq:identity_heisenberg} to the other. One extreme case would be the Mori projection operator \cite{Mori_1965} which projects onto only one observable, i.e.~the sum in eq.~\eqref{eq:general_projection} runs over only one function $\phi_1$. The Zwanzig projection operator is the opposite extreme case, because it requires a complete basis. It is this requirement that implicitly imposes the condition of minimal dissipation. To see this, we note that  the term {\it minimal} here refers to the norm induced by the inner product. Under this norm, 
\begin{equation}
\norm{\mathcal{P}B}\le\norm{\mathcal{P}^ZB} \;\forall B , \mathcal{P}\;,
\end{equation}
where $\mathcal{P}$ is any projection operator, $B$ is any observable and $\mathcal{P}^Z$ is the Zwanzig projection operator as defined in eq.~\eqref{eq:defZwanzig}. Thus, the projection operator we chose in sec.~\ref{sec:EoMDrift} maximizes the drift term (i.e.~the conservative term) and minimizes the rest (i.e.~the dissipative term). 

The crucial difference between sec.~\ref{sec:Zwanzig} and ref.~\cite{Colla_2021} is the definition of the inner product that induces the norm. The inner product employed in ref.~\cite{Hayden_2021} does not contain a weight, while the inner product required for the Hamiltonian of mean force to appear in eq.~\eqref{eq:densityDriftHamMeanForce} contains the equilibrium measure $\varrho_\beta$. As demonstrated in ref.~\cite{vomEnde_2024} a unique decomposition of a given generator can be achieved with respect to weighted scalar products. 
We thus agree with Colla and Breuer, one can use the generalized master equation and its unique splitting to define work. However, if the aim is a thermodynamic interpretation, we propose a different inner product, namely \eqref{eq:classical_scalar_product} with a canonical equilibrium distribution as weight. This yields the correct limit for systems coupled to a heat bath.

This leaves us with the last difference: quantum mechanics versus classical mechanics.

\section{Projection operator for quantum systems}
\label{sec:Qproj}
It is not straightforward to transfer the reasoning for classical systems to quantum systems in a universally valid manner. In this section we will briefly discuss why this transfer is difficult in general, and then show systems and environments with equilibrium states that factorize are an exception.

For the classical Zwanzig projector eq.~\eqref{eq:defZwanzig} we began the constructions with the introduction of the extended set of relevant variables $\{\psi_\alpha\}$.
For a single relevant quantum observable $A$ the projectors onto its eigenstates can play the same role. 

Suppose $A$ has a discrete spectrum
\begin{eqnarray}
    A = \sum_j a_j \Pi_j
\end{eqnarray}
where $\Pi_j = \sum_n \vert a_{j,n} \rangle \langle a_{j,n} \vert$ is the projection on the eigenstates of the observable belonging to the corresponding eigenvalue $a_j$. 
Then we can introduce the projector as
\begin{equation}
    \mathcal{P}B = \sum_j \frac{\tr(\varrho \Pi_j B)}{\tr(\varrho \Pi_j)} \Pi_j \label{eq:analogous_QMprojector},
\end{equation} 
and its adjoint 
\begin{equation}
    \mathcal{P}^\dagger \rho = \sum_j \frac{\tr(\rho \Pi_j)}{\tr(\varrho \Pi_j)}  \varrho \Pi_j
\end{equation}

Due to orthogonality, we have $\Pi_j\Pi_k = \delta_{jk}\Pi_j$, much the same as in eq.~\eqref{eq:classical_projections_psi}.
Both $\mathcal{P}$ and $\mathcal{P}^\dagger$ are idempotent. 
The projected $\mathcal{P}^\dagger \rho$ is still a density matrix. 
It can be interpreted as the projection on the equilibrium state after a non-selective measurement of $A$. 
By construction, the expectation value of $A$ is preserved, i.e., $\tr(A\rho )=\tr(A\mathcal{P}^\dagger \rho)$.

Although the projection operators are structurally very similar to the classical case, we cannot apply them correspondingly. 
As soon as we project onto a set of observables the order of the operators will matter. Suppose $A=\{A_k\}_k$ with respective decompositions $A_k = \sum_j a_{k,j} \Pi_{k,j} $. We cannot construct an operator $\Pi_j$ with the same properties as $\psi_\alpha= \prod_k \delta (A_k-\alpha_k)$ since for general $A_k$ the projectors will not commute. Or rephrased, it matters in which order the non-selective measurements are performed on the equilibrium state.

In addition, this construction does not produce an equation of motion for the reduced density matrix $\rho_S$ of a composite system. 
For a general density matrix, we find that $\rho_S = \tr_E(\rho)$ is not equal to $\tr_E(\mathcal{P}^\dagger \rho)$. Equality is only achieved if $\rho_S$ commutes with the relevant observable.

Instead, the projector can be built to project out all operators acting in $\mathcal{H}_S$. To implement this, we need a suitable scalar product similar to eq.~\eqref{eq:classical_scalar_product}. There is no unique quantum analogue, but an entire class of scalar products
\begin{equation}
    X,Y \mapsto \tr(\varrho_\beta^\alpha X^\dagger \varrho_\beta^{1-\alpha} Y), \qquad \alpha \in [0,1] \label{eq:deformed_scalar}
\end{equation}
which give a weight to the commutativity property with $H$ \cite{Rudoy_2009}.
A common choice in quantum statistical mechanics is to average over $\alpha$
\begin{equation}
    X,Y \mapsto (X,Y) = \int_0^1 \dif \alpha \, \tr \left( \varrho_\beta^\alpha X^\dagger \varrho_\beta^{1-\alpha} Y \right) \label{eq:scalar_product}.
\end{equation}
This scalar product is known under various names like Mori scalar product, Kubo's canonical correlation, Bogoliubov inner product or Duhamel two-point function \cite{Grabert_1977_equivalence, Grabert_1982, Petz_1993}. 

The corresponding similarity transformation to relate \eqref{eq:scalar_product} to the Hilbert-Schmidt product is given by
\begin{eqnarray}
    \Sigma X = \int_0^1 \dif \alpha \, \varrho_\beta^{\alpha} X \varrho_\beta^{1-\alpha}. \label{eq:Kubo-trafo}
\end{eqnarray}

The advantage of this transformation is, that
\begin{eqnarray}
    \frac{1}{\beta}\Sigma [X, \ln \varrho_\beta] = -\Sigma [X,  H ] = \Sigma \mathcal{L}X = \frac{1}{\beta}[X,\varrho_\beta] \label{eq:miracle_relation}
\end{eqnarray}
holds in analogy to the relation for the classical case, eq.~\eqref{eq:Poisson_relation}.

In ref.~\cite{Grabert_1982} this argument is used to define the projector and its adjoint as
\begin{eqnarray}
    \mathcal{P} X &=& \Sigma_S^{-1} \tr_E(\Sigma X) \\
    \mathcal{P}^\dagger \rho &=& \Sigma \Sigma_S^{-1} \tr_E(\rho)  
\end{eqnarray}
where $\Sigma_S$ maps system operators $X_S\in \mathcal{H}_S$ to system operators: $\Sigma_S X_S = \tr_E (\Sigma X_S)$.
With these definitions, we formally have the desired $\mathcal{P}X_S = X_S$ and $\tr_E(\mathcal{P}^\dagger \rho)=\tr_E(\rho)=\rho_S$, i.e.~all observables of the system are relevant and the relevant density yields the reduced system state if we trace over the environmental degrees of freedom.

We consider the drift term in the equation for the relevant observables and obtain
\begin{eqnarray}
    e^{i\mathcal{L}t} \mathcal{P} i\mathcal{L}X_S &=&  ie^{i\mathcal{L}t} \Sigma_S^{-1} \tr_E (\Sigma \mathcal{L} X_S) \\
    &=& ie^{i\mathcal{L}t } \Sigma_S^{-1} \tr_E (\frac{1}{\beta} [X_S,\varrho_\beta]) \\
    &=& ie^{i\mathcal{L}t} \Sigma_S^{-1} \frac{1}{\beta} [X_S,\varrho_{S,\beta}]. \label{eq:drift_obs_QM}
\end{eqnarray}

For the last step the relation~eq.~\eqref{eq:miracle_relation} has to hold when replacing $\Sigma \to \Sigma_S$ and $\varrho_\beta \to \varrho_{S,\beta}$, i.e.
\begin{equation}
    \frac{1}{\beta} \Sigma_S[X_S, \ln\varrho_{S,\beta}] = \frac{1}{\beta} [X_S,\varrho_{S,\beta}] \label{eq:necessary-cond}
\end{equation}
has to be fulfilled.
Under this condition the desired commutator with the Hamiltonian of mean force is obtained in eq.~\eqref{eq:drift_obs_QM} from $\frac{1}{\beta } [X_S,\rho_{S,\beta}]=-\Sigma_S[X_S,H_S^\ast]$.

This is a promising approach, but the validity of eq.~\eqref{eq:necessary-cond} depends on the structure of $\varrho_\beta$ and the type of correlations between the system of interest and the environment. If there are no correlations $\varrho_\beta = \varrho_{S,\beta} \otimes \varrho_{E,\beta}$ eq.~\eqref{eq:necessary-cond} indeed holds since
\begin{eqnarray}
    \Sigma_S [X_S,H_S^\ast] &=& \tr_E \int_0^1 \dif \alpha\; \varrho_{\beta}^\alpha [X_S\otimes \mathbb{I}_E,H_S^\ast \otimes \mathbb{I}_E] \varrho_{\beta}^{1-\alpha}  \\
    &=& \int_0^1 \dif \alpha \; \varrho_{S,\beta}^\alpha [X_S,H_S^\ast ] \varrho_{S,\beta}^{1-\alpha} \\
    &=& -\frac{1}{\beta}[X_S,\varrho_{S,\beta}] \; .
\end{eqnarray}

We leave the question whether eq.~\eqref{eq:necessary-cond} also holds if classical and even quantum correlations are involved for future investigation. The interested reader can refer to  apx.\ref{sec:Trafo_for_S} for an initial approach. 
Further, it remains to analyse whether the scalar product based on the transformation $\Sigma$ allows for a unique decomposition of the master equation \eqref{eq:TCL}. In ref.~\cite{vomEnde_2024} it was established for the deformed scalar product eq.~\eqref{eq:deformed_scalar} with $\alpha =\frac{1}{2}$ that such a splitting exists but a proof for arbitrary $\alpha$ requires different techniques.

\section{Conclusion}
We have suggested a definition of work performed on systems coupled to an environment, where the environment does not need to be in an equilibrium state and the coupling does not need to be weak. Our suggestion is based on projection operator techniques and the concept of minimal dissipation, which has recently been brought forward as a means to define work in open quantum systems \cite{Colla_2021,Colla_2025,Picatoste_2024}. For classical systems we show that the concept of minimal dissipation can be used to obtain a definition of work that has the correct limit for quasi-static processes performed on systems coupled to a heat bath. This is achieved by using an inner product with the global equilibrium distribution as weight. 
In contrast, the original proposal \cite{Colla_2021} established a splitting of the reduced dynamics based on an unweighted product and the minimization of the so-called dissipative part. 
Our investigation suggests that the corresponding effective Hamiltonian is not directly related to thermodynamic work. 

For systems and
environments with equilibrium states that factorize, the ideas can be transferred directly from the classical to the quantum mechanical case. For more complex cases there is no general procedure to define an appropriately weighted inner product and a projection operator. 
We assume that there is no unique definition.
Thus, the inner product has to be chosen case by case such that it is appropriate to a given system. 
The approach that we proposed admits a work definition in composite systems that equilibrate but is limited to situations where the Hamiltonian is time independent. As soon as the composite system is subject to external driving the drift term obtained with the Zwanzig projector does not admit a straightforward relation to a Hamiltonian of mean force \cite{Izvekov_2021,glatzel2022interplay}.  

We hope that our study will stimulate further investigation in this direction.

\begin{acknowledgments}
We thank Heinz-Peter Breuer and Anja Kuhnhold for helpful comments.
This project has received support by the DFG funded Research Training Group "Dynamics of Controlled Atomic and Molecular Systems" (RTG 2717).

\end{acknowledgments}

\section*{Data availability}
Data sharing does not apply to this article as no datasets were generated or analysed during
the current study.

\section*{Declarations}
\subsection*{Conflicts of interest}
The authors have no competing interests to declare that are relevant to the content of this article.

\begin{appendix}

\section{Notation}
In statistical mechanics, typically a different notation is used (see e.g. \cite{Evans_Morriss_2008}). Instead of indicating the Heisenberg picture by an index $H$, the argument is written as a function of time, i.e., $B_H(t) = B(\Gamma(t)) = B(\Gamma^t)$. The Liouvillian for the observable $\mathcal{L}_H(t) = \mathcal{L}(\Gamma(t),t)$ is there referred to as phase space- or p-Liouvillian, while the term f-Liouvillian is used for $\mathcal{L}$ which governs the evolution of the distribution function. In the same spirit the propagators which evolve the phase functions and distribution from the initial time to time $t$ are p- and f-propagators respectively.
The p-propagator can be defined as $\Gamma(t) = U_R(0,t)\Gamma(0)$.
With this notation the Liouville equation \eqref{eq:obs_Liouville} can also be expressed as 
\begin{eqnarray}
    \frac{\dif}{\dif t} B(\Gamma(t)) &=& \dot{\Gamma}(\Gamma(t),t) \left(\frac{\partial B(\Gamma)}{\partial \Gamma}\right)_{\Gamma=\Gamma(t)} \\
    &=& U_R(0,t) \dot{\Gamma}(\Gamma(0),t) \left(\frac{\partial B(\Gamma}{\partial \Gamma}\right)_{\Gamma=\Gamma(0)} \\
    &=& U_R(0,t) i\mathcal{L}(\Gamma(0),t) B(\Gamma(0)) \\
    &=& \frac{\partial}{\partial t} U_R(0,t) B(\Gamma(0))
\end{eqnarray}
which yields an operator equation for $U_R$.
This is formally solved by 
\begin{widetext}    
\begin{eqnarray}
    U_R(t',t)&=&\exp_R \left(\int_{t'}^t \dif s \, i \mathcal{L}(\Gamma(t'),s)\right) \\
    &=& 1+ \sum_{n=1}^\infty \int_{t'}^t \dif s_1  \int_{t'}^{s_1} \dif s_2 \dots \int_{t'}^{s_{n-1}} \dif s_n i\mathcal{L}(\Gamma(t'),s_n) \dots i\mathcal{L}(\Gamma(t'),s_1)
\end{eqnarray}
\end{widetext}
which is the right time-ordered exponential. The Liouvillians act on the phase function in an anticausal order \cite{Evans_Morriss_2008}. For $t'=0$ we deal with the Schrödinger picture Liouvillian. It can be shown that $U_R(0,t)$ is equal to a left time-ordered exponential with causal ordering of Heisenberg Liouvillians \cite{teVrugt_2019}
\begin{eqnarray}
    \exp_R \left( i\int_0^t \dif s \mathcal{L}(s)\right) &=& \exp_L\left(i \int_0^t \dif s \mathcal{L}_H(s)\right) \\
    \exp_L\left(-i\int_0^t \dif s \mathcal{L}(s)\right) &=& \exp_R\left(-i\int_0^t \dif s \mathcal{L}_H(s)\right)
\end{eqnarray}  

\subsection{Operator ordering}
The Heisenberg picture is especially convenient to determine correlations of observables $B_i, \ i=1, \dots, n$ at different times $t_i$ of the form
\begin{equation}
    \langle B_1(t_1)\dots B_n (t_n) \rangle = \tr (B_1(t_1) \dots B_n (t_n) \rho(0)\rangle   
\end{equation}
Note that the functions under the classical phase-space integral can be permuted arbitrarily. But the quantum mechanical trace is only invariant under cyclic permutations. Accordingly there are different possible multi-time expectations of Heisenberg operators that are reduced to the same correlation in the classical limit.

\section{Propagator decomposition}
\label{apx:decomposition}
The identity \eqref{eq:identity_heisenberg} can be confirmed by differentiation or motivated by the physical interpretation as discussed in ref.~\cite{Grabert_1982}. The propagator can be decomposed in a sum by inserting the identity $e^{i\mathcal{L}t}=e^{i\mathcal{L}t}(\mathcal{P}+\mathcal{Q})$. Applying the first term to an arbitrary observable $B$ yields a linear combination of relevant variables. The aim is to find an expression for the other, orthogonal part in terms of the information about the relevant dynamics. 
Taking the partial time derivative we have
\begin{eqnarray}
    \frac{\partial}{\partial t} e^{i\mathcal{L}t}\mathcal{Q} &=& e^{i\mathcal{L}t}i\mathcal{L}\mathcal{Q} \\
    &=&e^{i\mathcal{L}t}\mathcal{P} i\mathcal{L}\mathcal{Q} + e^{i\mathcal{L}t}\mathcal{Q} i\mathcal{L}\mathcal{Q}.
\end{eqnarray}
In the second step the identity is inserted again, and we obtain an inhomogeneous equation for $e^{i\mathcal{L}t}\mathcal{Q}$. The inhomogeneous term is a linear combination of the variables of interest. 
The solution to the homogeneous part of the equation is $e^{i\mathcal{L}s}\mathcal{Q}G(s,t)$ and thus 
\begin{equation}
    e^{i\mathcal{L} t}\mathcal{Q} =  e^{i\mathcal{L} s}\mathcal{Q}G(s,t) + \int_s^t \dif t' e^{i\mathcal{L}t'} \mathcal{P}i\mathcal{L} \mathcal{Q}G(t',t) .
\end{equation}

With this we obtain eq.~\eqref{eq:identity_heisenberg}. 

\section{Similarity Transformation and Kubo-relation}
\label{sec:Trafo_for_S}
\begin{widetext}

It remains to investigate under which more general conditions the equality
\begin{eqnarray}
    -\Sigma_S[X_S,H_S^\ast]=\frac{1}{\beta } [X_S,\rho_{S,\beta}]
\end{eqnarray} 
holds. According to the Kubo-relation \cite{Wilcox_1967}
\begin{eqnarray}
    -\int_0^1 \dif \alpha\ \varrho_{S,\beta}^\alpha [X_S,H_S^\ast] \varrho_{S,\beta}^{1-\alpha} &=& \frac{1}{\beta } [X_S,\varrho_{S,\beta}] \qquad \text{where } \varrho_{S,\beta} \propto e^{-\beta H_S^\ast}\\
\end{eqnarray}

We continue the discussion from sec.~\ref{sec:Qproj} and suppose that the global equilibrium state is a classical state with respect to local measurements, then it can be represented as \cite{Luo_2008}
\begin{eqnarray}
    \varrho_\beta = \sum_{ij} p_{ij} P_i \otimes Q_j \quad \Rightarrow \quad \varrho_\beta^\alpha &=& \sum_{ij} p_{ij}^\alpha P_i \otimes Q_j
\end{eqnarray}
where $P_i$ and $Q_j$ are projectors onto some orthonormal basis in $\mathcal{H}_S$ and $\mathcal{H}_E$ respectively. In fact, these are then the spectral projectors of the reduced states ($\varrho_{S,\beta} = \sum_i p_i P_i$, $\sum_j p_{ij} = p_i$ and $\varrho_{E,\beta} = \sum_j q_j Q_j$, $\sum_ip_{ij}=q_j$). 
This class of states includes non-trivial combinations of product states if they commute. 

\begin{eqnarray}
     -\tr_E \int_0^1 \dif \alpha \;\varrho_{\beta}^\alpha [X_S\otimes \mathbb{I}_E,H_S^\ast \otimes \mathbb{I}_E] \varrho_{\beta}^{1-\alpha} &=& -\tr_E \int_0^1 \dif \alpha \sum_{ij} p_{ij}^\alpha P_i \otimes Q_j [X_S\otimes \mathbb{I}_E,H_S^\ast \otimes \mathbb{I}_E] \sum_{nm} p_{nm}^{1-\alpha}P_n \otimes Q_m \\
     &=& -\tr_E \int_0^1 \dif \alpha \sum_{ij} p_{ij}^\alpha P_i [X_S,H_S^\ast ] \sum_{nm} p_{nm}^{1-\alpha}P_n \otimes Q_j Q_m \\
    &=& - \int_0^1 \dif \alpha \sum_{ij} p_{ij}^\alpha P_i [X_S,H_S^\ast ] \sum_{nm} p_{nm}^{1-\alpha}P_n \delta_{jm} \\
    &=& - \int_0^1 \dif \alpha \sum_{ij} p_{ij}^\alpha P_i [X_S,\sum_k h_k P_k ] \sum_{n} p_{nj}^{1-\alpha}P_n \\
    &=& - \int_0^1 \dif \alpha \sum_{ijkn} \left(p_{ij}^\alpha P_i X_S h_k P_k  p_{nj}^{1-\alpha} \delta_{kn} - p_{ij}^\alpha h_k P_k \delta_{ki} X_S p_{nj}^{1-\alpha}P_n\right) \\
    &=&- \int_0^1 \dif \alpha\left(\sum_{ijk} p_{ij}^\alpha P_i X_S h_k P_k  p_{kj}^{1-\alpha}  - \sum_{jkn}p_{kj}^\alpha  h_k P_k  X_S p_{nj}^{1-\alpha}P_n\right) \label{eq:q1}
\end{eqnarray}
and
\begin{eqnarray}
     -\int_0^1 \dif \alpha \varrho_{S,\beta}^\alpha [X_S,H_S^\ast] \varrho_{S,\beta}^{1-\alpha} &=& -\int_0^1 \dif \alpha \sum_i p_i^\alpha P_i [X_S, \sum_k h_k P_k ] \sum_n p_n^{1-\alpha} P_n \\
     &=&  -\int_0^1 \dif \alpha \sum_{ikn} \left( p_i^\alpha P_i X_S  h_k P_k  p_n^{1-\alpha} \delta_{kn} - p_i^\alpha \delta_{ik} h_k P_k X_S    p_n^{1-\alpha} P_n\right)\\
    &=& -\int_0^1 \dif \alpha \left(\sum_{ik} p_i^\alpha P_i X_S  h_k P_k  p_k^{1-\alpha}- \sum_{kn}p_k^\alpha  h_k P_k X_S    p_n^{1-\alpha} P_n\right) \label{eq:Q2}
\end{eqnarray}

The spectral decomposition for the mean force Hamiltonian includes the same projectors and could be expressed as $H_S^\ast = \sum_k h_k P_k$.
Now, is \eqref{eq:q1} equal to \eqref{eq:Q2}?

If this is indeed true, the next step would be the study of separable equilibrium states, i.e.~ those that can be represented as a sum over product states $\varrho_\beta = \sum_j p_j\varrho_{S,j}\otimes \varrho_{E,j}$.
\end{widetext}

\end{appendix}

\bibliography{projector}

\end{document}